\documentclass[a4paper,12pt]{article}
\usepackage{graphicx}
\usepackage{epsfig}
\begin{document}
\title{Young neutron stars with soft gamma ray emission and anomalous X-ray pulsar}

\author{Gennady S. Bisnovatyi-Kogan \\
Space Research Institute of Russian Academy of Sciences,\\ Profsoyuznaya 84/32, Moscow 117997, Russia,\\ and\\
National Research Nuclear University MEPhI \\ (Moscow Engineering Physics Institute),\\ Kashirskoe Shosse 31, Moscow 115409, Russia\\
Email: gkogan@iki.rssi.ru}
\date{}

\maketitle

\begin{abstract}
The observational properties of Soft Gamma Repeaters and Ano\-malous X-ray Pulsars (SGR/AXP) indicate to necessity of the energy source different from a rotational energy of  a neutron star. The model, where the source of the energy is connected with a magnetic field dissipation in a highly magnetized neutron star (magnetar) is analyzed. Some observational inconsistencies  are indicated for this interpretation.
The alternative energy source, connected with the nuclear energy of superheavy nuclei stored in the nonequilibrium layer of low mass neutron star is discussed.
\end{abstract}

\maketitle

\section{\label{sec:level1} Introduction}

Neutron stars (NS) are formed as a result of a collapse of the core of a massive star with a mass $M>\sim 12 M_\odot$.
Conservation of the magnetic flux gives an estimation of NS magnetic field as
 $B_{ns}=B_s (R_s  /R_{ns}   )^2$,
$B_{s}=10 \div 100$ Gs, at   $R \sim (3 \div 10) R_\odot,\,\,  R_{ns}   =10$ km,
$B_{ns} = 4\cdot 10^{11} \div 5\cdot 10^{13}$   Gs (Ginzburg 1964).

Estimation of the NS magnetic field is obtained in radio pulsars by measurements of
their rotational period and its time derivative, in the model of a dipole radiation, or
pulsar wind model, as ($E$, $I$, and $\Omega$  are NS rotational energy,  moment of inertia, and
rotational angular velocity,  respectively):

\begin{equation}
\label{magf}
 E_{rot} = 0.5 I\Omega^2, \quad \dot E_{rot} = AB^2\Omega^4,\quad
B  =  IP\dot P/4A \pi^2, \quad A=R^6/6c^3,
\end{equation}
$B$ is NS surface dipole magnetic field at its magnetic pole.
Timing observations of single radiopulsars (the rapidly rotating ones  connected with young supernovae remnants are
marked by star)
give the following estimation   $B_{ns}  =   2 \cdot 10^{11} \div 5\cdot 10^{13}$ Gs (Lorimer 2005).

\begin{figure}[htp]
\centerline{\hbox{\includegraphics[width=1.\textwidth,angle=-90]{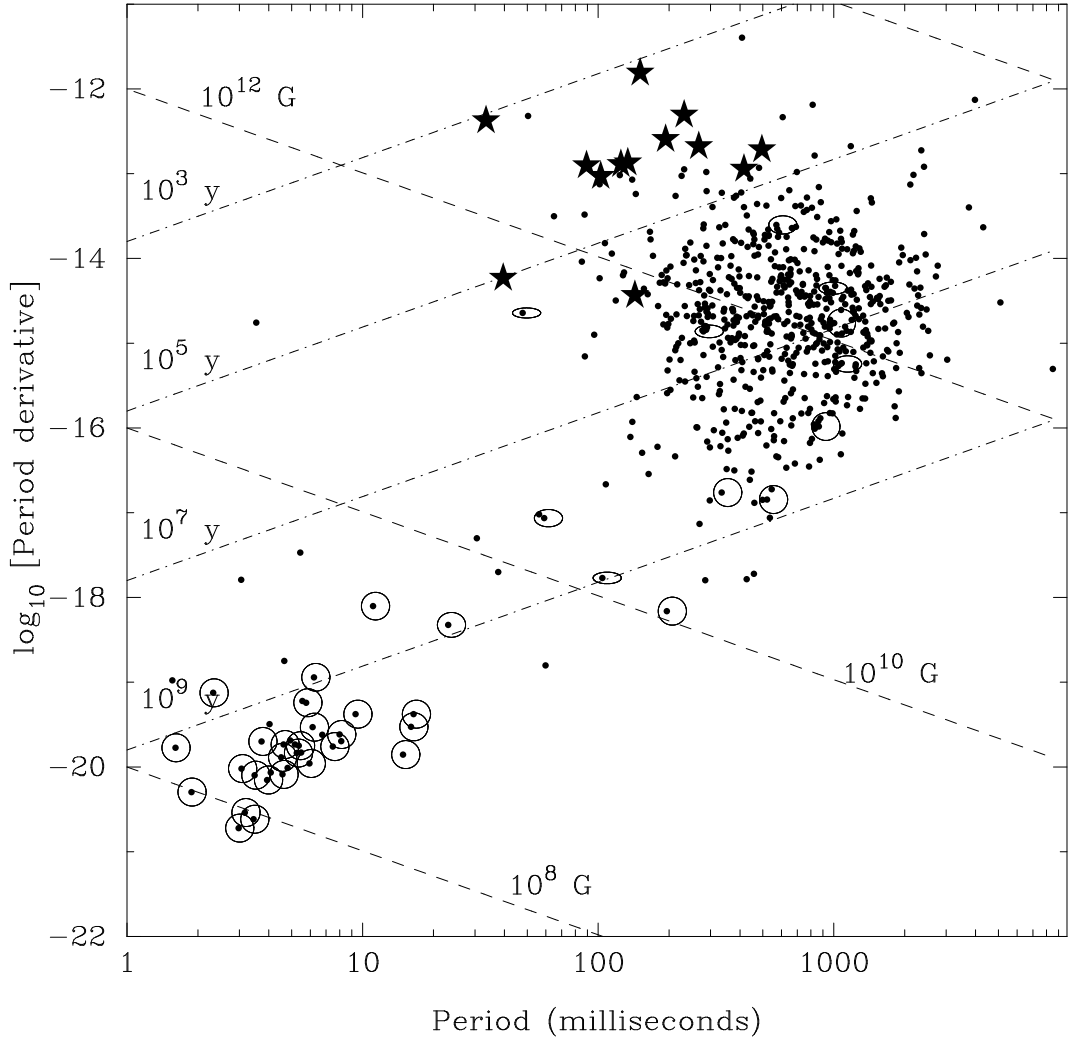}}}
\caption{$P$ - $\dot P$ diagram for radiopulsars. Pulsars in binary systems with low-eccentricity orbits are
encircled, and in high-eccentricity orbits are marked with ellipses. Stars
show pulsars suspected to be connected with supernova remnants,
 from Lorimer (2005).}
\label{fig1}
\end{figure}
\noindent The pulsars with a small magnetic field in the left lower angle decrease their magnetic field
during recycling by accretion in a close binary.  see Bisnovatyi-Kogan (2006).

SGR are single neutron stars with periods $2 \div 8$ seconds
They produce "giant bursts", when their luminosity $L$ in the peak increase $5 \div 6$ orders of magnitude.
Having a slow rotation, and small rotational energy, their observed average luminosity
exceeds rotational loss of energy more than 10 times, and orders of
magnitude during the giant outbursts.

It was suggested by Duncan and Thompson (1995), that the source of energy is their huge magnetic field,
2 or 3 order of  a magnitude larger, then the average field in radiopulsars. Such objects were called  magnetars.

\section{SGR,  giant bursts, and short GRB}

First two Soft Gamma Repeaters (SGR) had been discovered by KONUS
group in 1979. The first one, FXP 0520 - 66, was
discovered after the famous giant 5 March 1979 burst
(Mazets et al. 1979b,c; Golenetskii et al. 1979), see also
Mazets et.al (1982). In another
source B1900+14 only small recurrent bursts had been observed (Mazets et al. 1979a). Now
these sources are known under names SGR 0520 - 66 and SGR 1900+14
respectively. The third SGR 1806-20 was identified as a repetitive
source by Laros et.al. (1986a,b). The first detection of this source as
GRB070179 was reported by Mazets et al.(1981), and it was indicated by
Mazets et al. (1982), that this source, having an unusually soft spectrum,
can belong to a separate class of repetitive GRB, similar to
FXP0520 - 66 and B1900+14. This suggestion was completely confirmed.
The forth known SRG1627-41, showing giant burst, was discovered in
1998 almost simultaneously by BATSE (Kouveliotou et al. 1998a), and BeppoSAX
(Feroci et al. 1998). The giant bursts had been observed until now in 4 sources.

\subsection{SGR0526-66}

It was discovered due to a giant burst of 5 March 1979, projected to
the edge of the SNR N49 in LMC, and described by (Mazets et al. 1979b,c; Golenetskii et al. 1979, Mazets et.al 1982). Accepting the distance 55 kpc to LMC, the peak luminosity in the region $E_{\gamma}>30$ keV was
$L_p \ge 3.6\times 10^{45}$ ergs/s, the total energy release in the peak
$Q_p \ge 1.6 \times 10^{44}$ ergs, in the subsequent tail $Q_t=3.6
\times 10^{44}$ ergs. The short recurrent bursts have peak
luminosities in this region $L_p^{rec}=3\times 10^{41}\,-\, 3 \times
10^{42}$ ergs/s, and energy release $Q^{rec}=5\times 10^{40}\,-\, 7
\times 10^{42}$ ergs. The tail was observed about 3 minutes and had
regular pulsations with the period $P\approx 8$ s. There was not a
chance to measure $\dot P$ in this object.

\subsection{SGR1900+14}

\begin{figure}[htp]
\centerline{\hbox{\includegraphics[width=0.8\textwidth]{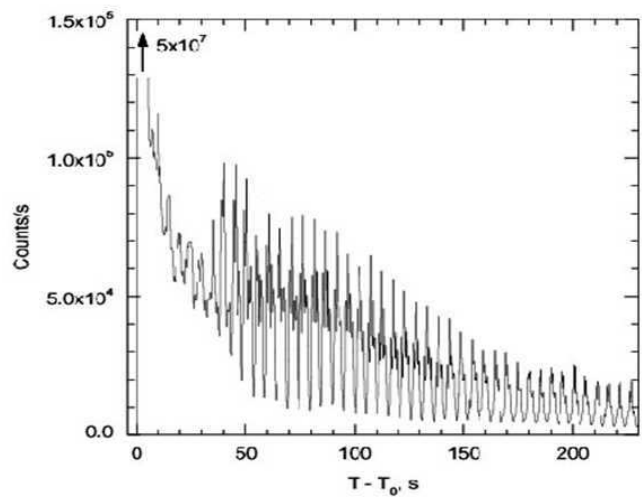}}}
\caption{The giant 1998 August 27 outburst of the soft gamma repeater SGR1900 + 14.  Intensity of the E > 15 keV radiation in presented, from Mazets et al. (1999c).} \label{fig3}
\end{figure}

Detailed observations of this source are described by
Mazets et al. (1999b,c), Kouveliotou et al. (1999), Woods et al. (1999).
 The giant burst was observed 27
August, 1998. The source lies close to the less than $10^4$ year old
SNR G42.8+0.6, situated at distance $\sim 10$ kpc. Pulsations had
been observed in the giant burst, as well as in the X-ray emission
observed in this source in quiescence by RXTE and ASCA. $\dot P$ was
measured, being strongly variable. Accepting the distance 10 kpc,
this source had in the region $E_{\gamma}>15$ keV: $L_p > 3.7\times
10^{44}$ ergs/s, $Q_p > 6.8\times 10^{43}$ ergs, $Q_t=5.2 \times
10^{43}$ ergs, $L_p^{rec}=2\times 10^{40}\,-\, 4\times 10^{41}$
ergs/s, $Q^{rec}=2\times 10^{39}\,-\, 6\times 10^{41}$ ergs,
$P=5.16$ s, $\dot P=5 \times 10^{-11}\,-\, 1.5\times 10^{-10}$ s/s.
The X-ray pulsar in the error box of this source was discovered by Hurley et al. (1999b).
This source was discovered also in radio band, at frequency 111 MHz as a faint,
$L_r^{max}=50$ mJy, radiopulsar (Shitov 1999), with the same $P$ and
variable $\dot P$, good corresponding to X-ray and gamma-ray
observations. The values of $P$ and average $\dot P$ correspond to
the rate of a loss of rotational energy $\dot E_{rot}=3.5 \times
10^{34}$ ergs/s, and magnetic field $B=8 \times 10^{14}$ Gs. The age
of the pulsar estimated as $\tau_p=P/2\dot P=700$ years is much less
than the estimated age of the close nearby SNR. Note that the observed X-ray
luminosity of this object $L_x=2\times 10^{35}\,\, - \,\, 2\times
10^{36}$ ergs/s is much higher, than rate of a loss of rotational
energy, what means that rotation cannot be a source of energy in
these objects. It was suggested that the main source of energy comes
from a magnetic field annihilation, and such objects had been called
as magnetars by Duncan and Thompson (1992). The light curve of the giant burst is given in Fig.\ref{fig3}.

\subsection{SGR1806-20}

The giant burst from this source was observed in December 27, 2004
(Palmer et al. 2005, Mazets et al. 2005, Frederiks et al. 2007).
 Recurrent bursts had been studied by Kouveliotou et al. (1998b), Hurley et al. (1999a). Connection with the Galactic radio SNR G10.0-03
was found. The source has a small but significant displacement from
that of the non-thermal core of this SNR. The distance to SNR is
estimated as 14.5 kpc. The X-ray source observed by ASCA and RXTE in
this object shows regular pulsations with a period $P=7.47$ s, and
average $\dot P=8.3\times 10^{-11}$ s/s. As in the previous case, it
leads to the pulsar age $\tau_p \sim 1500$ years, much smaller that
the age of SNR, estimated by $10^4$ years. These values of $P$ and
$\dot P$ correspond to $B=8\times 10^{14}$ Gs. $\dot P$ is not
constant, uniform set of observations by RXTE gave much smaller and
less definite value $\dot P=2.8(1.4)\times 10^{-11}$ s/s, the value
in brackets gives 1$\sigma$ error. The peak luminosity in the burst
reaches $L_p^{rec}\sim 10^{41}$ ergs/s in the region 25-60 keV, the
X-ray luminosity in 2-10 keV band is $L_x\approx 2\times 10^{35}$
ergs/s is also much higher than the rate of the loss of rotational
energy (for average $\dot P$) $\dot E_{rot} \approx 10^{33}$ ergs/s.

 The burst of December 27, 2004 in SGR 1806-20 was the greatest flare, $\sim 100$ times brighter than ever. It was detected by many satellites: Swift, RHESSI, Konus-Wind, Coronas-F, Integral, HEND et al..

\begin{figure}[htp]
\centerline{\hbox{\includegraphics[width=0.8\textwidth]{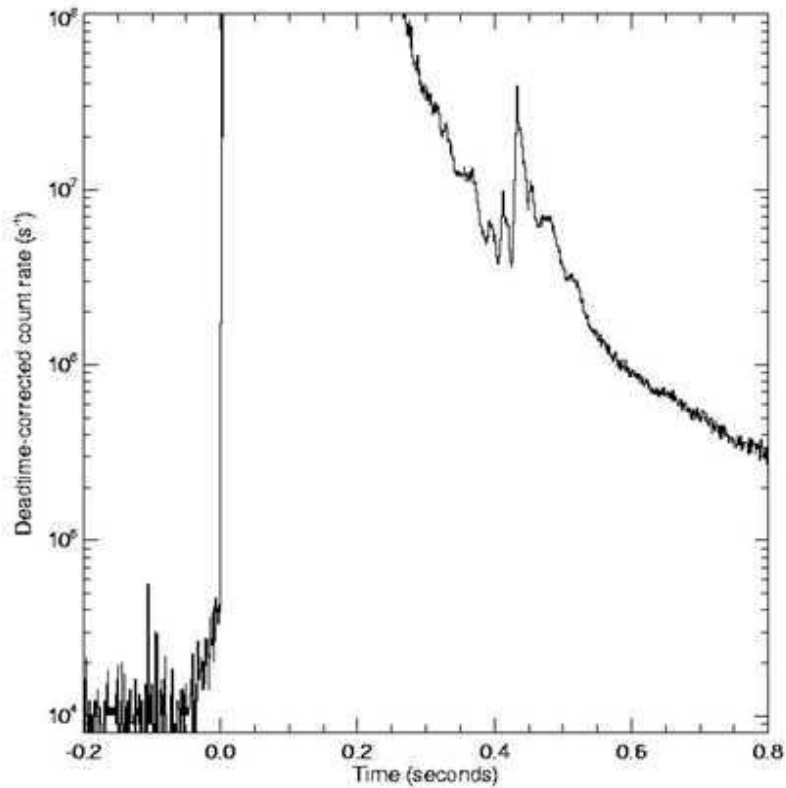}}}
\caption{SWIFT light curve of 27 December, 2004 giant burst in  SGR1806, from Palmer et al. (2005)} \label{fig4}
\end{figure}
\begin{figure}[htp]
\centerline{\hbox{\includegraphics[width=0.8\textwidth]{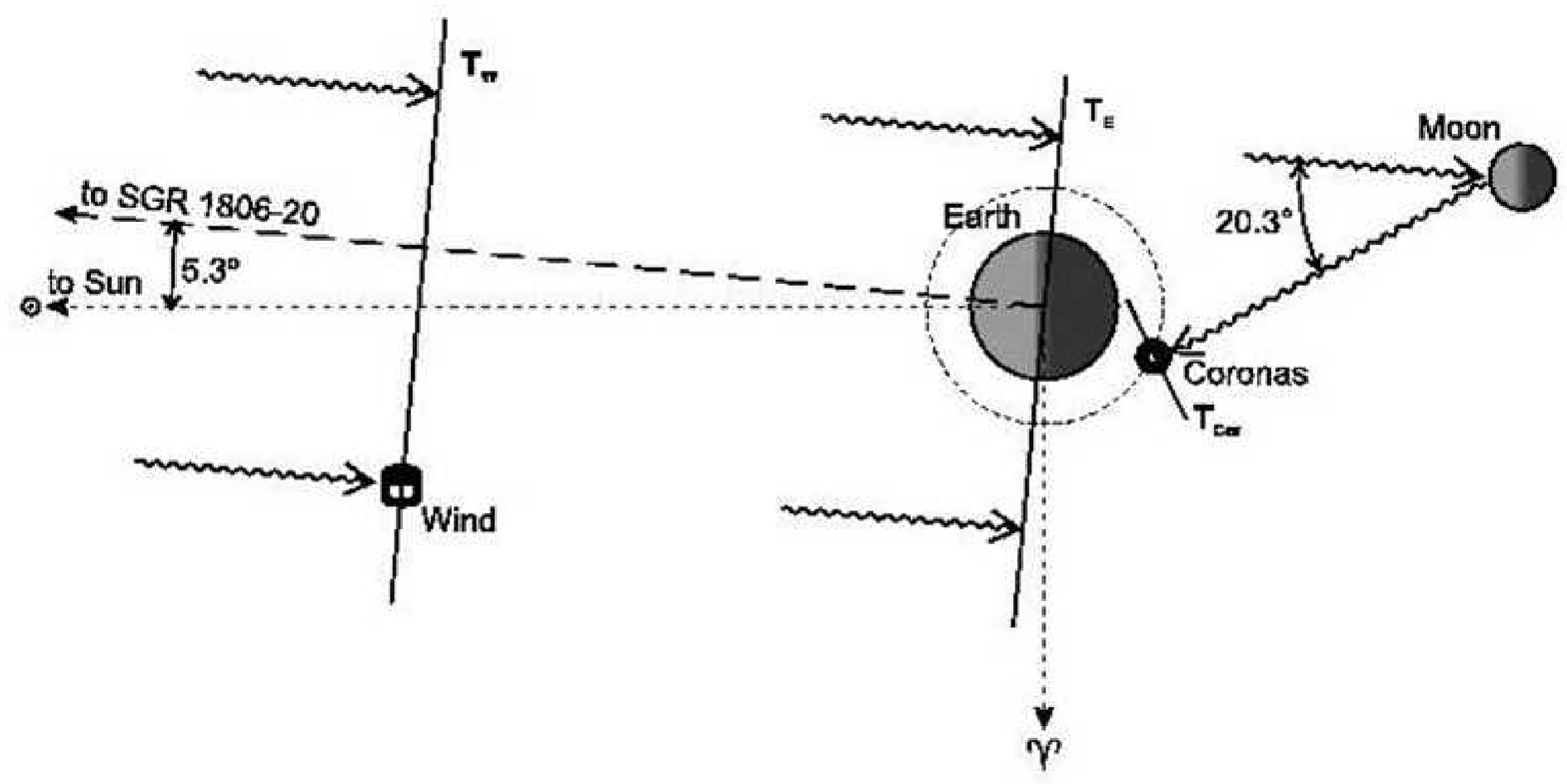}}}
\caption{The position of satellites Wind and Coronas-F relative to the Earth and Moon during the outburst, from Mazets et al. (2005), Frederiks et al. (2007b).} \label{fig5}.
\end{figure}
\begin{figure}[htp]
\centerline{\hbox{\includegraphics[width=0.8\textwidth]{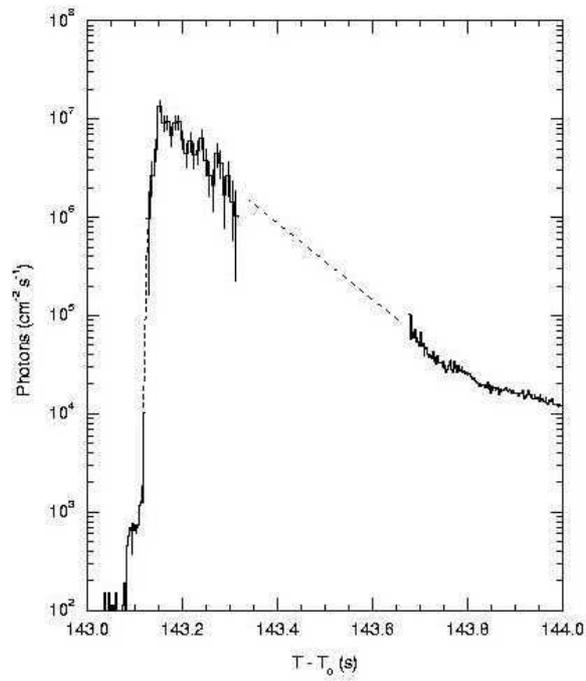}}}
\caption{ Reconstructed time history of the initial pulse. The upper part of the graph is derived from Helicon data while the lowerpart represents the Konus-Wind data. The dashed lines indicate intervals where the outburst intensity still saturates the Konus-Wind detector, but is not high enough to be seen by the Helicon, from Mazets et al. (2005), Frederiks et al. (2007b).} \label{fig5a}
\end{figure}

Very strong luminosity of this outburst permitted to observe the signal, reflected from the moon by the HELICON instrument on the board of the satellite Coronas-F. The position of satellites Wind and Coronas-F relative to the Earth and Moon during the outburst are given in Fig.\ref{fig5}, and the reconstructed full light curve of the outburst is given in Fig.\ref{fig5a} from Mazets et al. (2005), Frederiks et al. (2007a).

\subsection{SRG1627-41}

Here the giant burst was observed 18 June 1998, in addition to
numerous soft recurrent bursts. Its position coincides with the SNR
G337.0-0.1, assuming 5.8 kpc distance. Some evidences was obtained
for a possible periodicity of 6.7 s, but giant burst did not show
any periodic signal (Mazets et al. 1999a), contrary to three other giant burst
in SGR. The following characteristics had been observed with a time
resolution 2 ms at photon energy $E_{\gamma}> 15$ keV:
 $L_p \sim 8\times 10^{43}$ ergs/s,
$Q_p \sim 3\times 10^{42}$ ergs, no tail of the giant burst had been
observed. $L_p^{rec}=4\times 10^{40}\,-\, 4\times 10^{41}$ ergs/s,
$Q^{rec}=10^{39}\,-\, 3\times 10^{40}$ ergs. Periodicity in this
source is not certain, so there is no $\dot P$.

\subsection{SRG giant bursts in other galaxies}

The similarity between giant bursts in SGR, and short GRB was noticed by
by Mazets et al. (1999), Bisnovatyi-Kogan (1999).
The experiment KONUS-WIND had observed two short GRB, interpreted as giant bursts of SGR.
The first one, GRB070201, was observed
in M31 (Andromeda),  1 February, 2007.
The energy of the burst is equal to $~ 1 \cdot E^{45}$ erg, in consistence with
giant bursts of other SGR   Mazets et al. (2008).
The second short burst, GRB051103, was observed in the galaxy M81,  3 November 2005.
The energy of the burst is equal  to $~ 7\cdot E^{46}$ erg
(Golenetskii et al. 2005, Frederiks et al. 2007).

\section{Estimations of the magnetic fields\\ in SGR/AXP}

\begin{figure}[htp]
\centerline{\hbox{\includegraphics[width=0.8\textwidth]{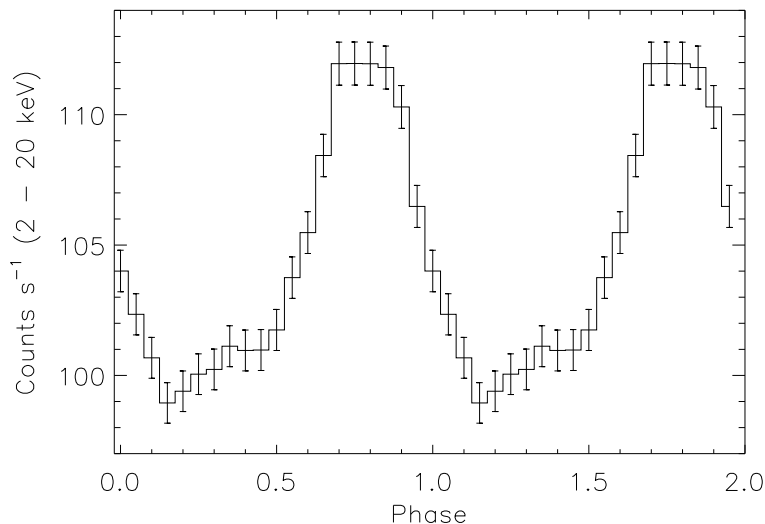}}}
\caption{The epoch folded pulse profile of SGR 1900 + 14 (2-20 keV) for the May 1998 RXTE observations, from Kouveliotou et al. (1999).}
\label{fig6}
\end{figure}

\begin{figure}[htp]
\centerline{\hbox{\includegraphics[width=0.8\textwidth]{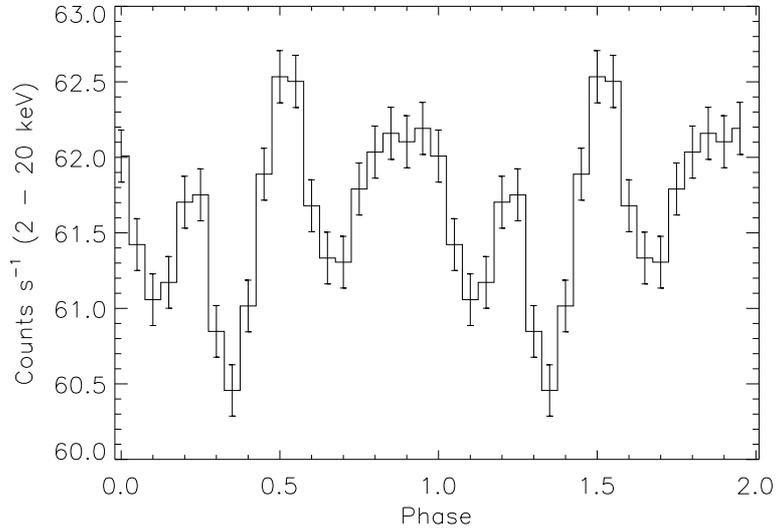}}}
\caption{The epoch folded pulse profile of SGR 1900 + 14 (2-20 keV) for the August 28, 1998 RXTE observation. The plot is exhibiting two phase cycles, from Kouveliotou et al. (1999).}
\label{fig7}
\end{figure}

\begin{figure}[htp]
\centerline{\hbox{\includegraphics[width=0.8\textwidth]{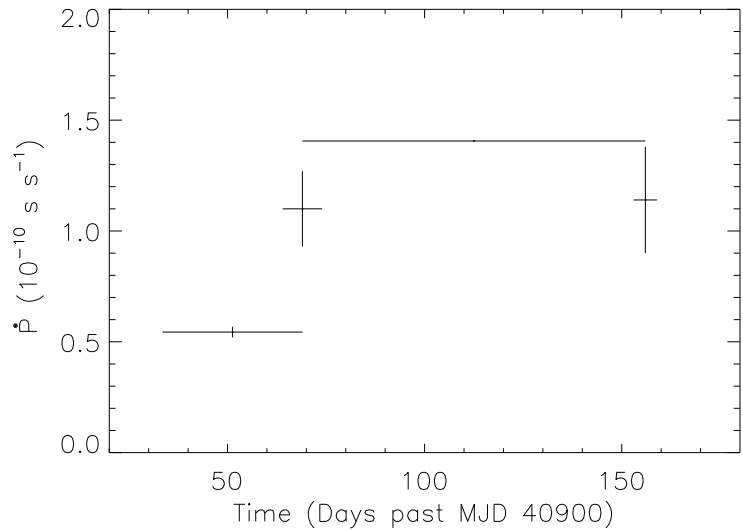}}}
\caption{The evolution of "Period derivative" versus time since the first period measurement of SGR 1900+14 with ASCA in \cite{hur99a}.
The time is given in Modified Julian Days (MJDs), from Kouveliotou et al. (1999)} \label{fig8}
\end{figure}

Despite the fact, that rotation energy losses are much smaller than the observed luminosity, for estimation of the magnetic field strength in these objects used the same procedure as in radio pulsars, based on measurements of $P$ and $\dot P$, and using (\ref{magf}). The first measurements have been done for  SGR 1900 + 14, in different epochs by measurements of satellites RXTE and ASCA
(Kouveliotou et al. 1999), presented in Figs.\ref{fig6}-\ref{fig8}.

\noindent The pulse shape is changing from one epoch to another, inducing errors in finding derivative of the period. The big jump in $\dot P$, visible in Fig \ref{fig8} looks out surprising. for magnetic dipole losses, because it needs a considerable jump in the magnetic field strength, prohibited by self induction effects. Contrary, in the model of pulsar wind rotational energy losses it looks quite reasonable, that these losses strongly increase during the giant burst, when the $\dot P$ jump was observed.

 Another evidence in favour of the magnetar magnetic field was connected with the absorption lines in the spectrum of SGR 1806-20, observed by RXTE in November 1996 (Ibrahim et al. 2002). The main line corresponds to magnetic field $(5\div 7)\cdot
10^{11}$ Gs, when interpreted as an electron cyclotron line. In order to preserve the magnetar model, the authors \cite{ibr02} suggested that this line is connected with the proton motion, increasing the magnetic field estimation almost 2000 times. It is connected, however, with a drastic, $\sim 4\cdot 10^6$, decrease in the absorption cross-section, compared to the electron cyclotron line. Therefore, if this cyclotron line is real, its connection with the proton is very improbable.

\begin{figure}[htp]
\centerline{\hbox{\includegraphics[width=0.8\textwidth]{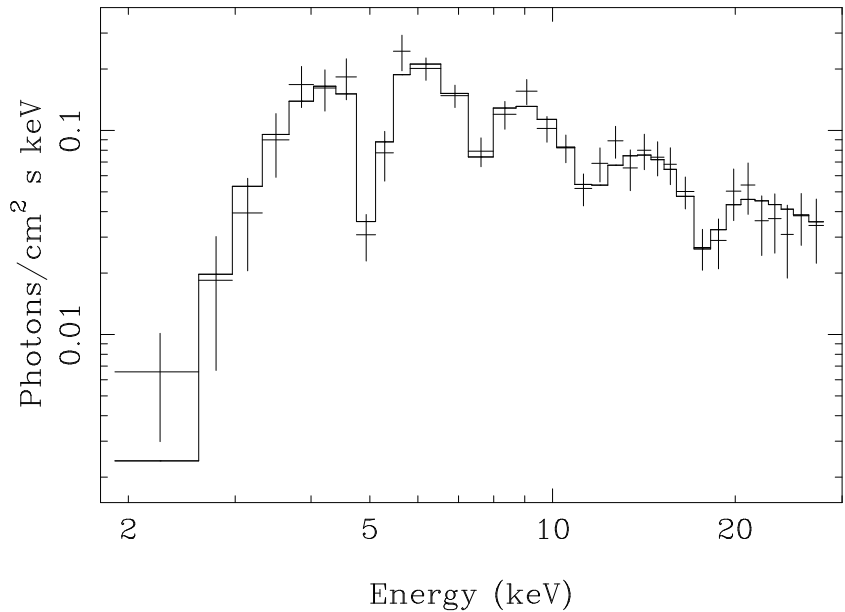}}}
\caption{SGR 1806-20
spectrum and best-fit continuum model for the second precursor interval with 4 absorption lines (RXTE/PCA 2-30 keV), from Ibrahim et al. (2002)} \label{fig9}
\end{figure}

\section{Radiopulsars with very high magnetic \\ fields and slow rotation}
\label{radio}

Radio pulsars are rotating neutron stars that emit beams of radio waves from regions above their magnetic poles. Popular theories of the emission mechanism require continuous electron-positron pair production, with the potential responsible for accelerating the particles being inversely related to the spin period. Pair production will stop when the potential drops below a threshold, so the models predict that radio emission will cease when the period exceeds a value that depends on the magnetic field strength and configuration. It was shown by Young et al. (1999a,b)
 that the pulsar J2144-3933, previously thought to have a period of 2.84s, actually has a period of 8.51s, which is by far the longest of any known radio pulsar. Moreover, under the usual model assumptions, based on the neutron-star equations of state, this slowly rotating pulsar should not be emitting a radio beam. Therefore either the model assumptions are wrong, or current theories of radio emission must be revised.
The period 8.51 second is characteristic for SGR/AXP objects, but this pulsar does not show any violent behaviour, and behave like ordinary radio pulsar.

Soon after this discovery, several other  radio  pulsars were found, where also $\dot P$, and therefore magnetic field strength was measured
(Manchester et al. 2001, Camilo et al. 2000, McLaughlin et al. 2003,2004). These pulsars include:

 1. PSR J1119 - 6127,  P = 0.407 s,  $\dot P = 4.0 \cdot 10^{-12}$ s/s, $B=4.1\cdot 10^{13}$ G;

 2. PSR J1814 - 1744, P = 3.975 s,  $\dot P = 7.4 \cdot 10^{-13}$ s/s, $B = 5.5 \cdot 10^{13}$ G;

\noindent It was noted by Camilo et al. (2000), that "Both PSR J1119 — 6127 and PSR J1814 —1744 show apparently normal radio emission in a regime of magnetic field strength where some models predict that no emission should occur. Also, PSR J1814 —1744 has spin parameters similar to the anomalous X-ray pulsar (AXP) IE 2259 + 586, but shows no discernible X-ray emission. If AXPs are isolated, high magnetic field neutron stars (“ magnetars ”), these results suggest that their unusual attributes are unlikely to be merely a consequence of their very high inferred magnetic fields."

 3.  PSR J1847 - 0130, P=6.7 s, $\dot P = 1.3 \cdot 10^{-12}$ s/s, B=$9.4 \cdot 10^{13}$ G.

 \noindent It was noted in the paper of McLaughlin et al. (2003), with the title "PSR J1847—0130: A RADIO PULSAR WITH MAGNETAR SPIN CHARACTERISTICS",  that "The properties of this pulsar prove that inferred dipolar magnetic field strength and period cannot alone be responsible for the unusual high-energy properties of the magnetars and create new challenges for understanding the possible relationship between these two manifestations of young neutron stars."

 4. PSR J1718 - 37184, P= 3.4 s , \qquad  $B= 7.4 \cdot 10^{13}$ G.

 \noindent It was noted in the paper of McLaughlin et al. (2004), that "These fields are similar to those of the anomalous X-ray pulsars (AXPs), which growing evidence suggests are “magnetars”. The lack of AXP-like X-ray emission from these radio pulsars (and the non-detection of radio emission from the AXPs) creates new challenges for understanding pulsar emission physics and the relationship between these classes of apparently young neutron stars."

\section{SGR/AXP with low magnetic fields and moderate rotation}
\label{small}

SGR/AXP  J1550-5418 (1E 1547.0-5408) was visible in radio band, showing pulsations with a period $P=2.069$ s )Camilo et al. 2007). The pulsations with the same period  have been observed first only in the soft X ray band by XMM-Newton (Halpern et al. 2008). In the hard X ray region statistics of photons was not enough for detection of pulsations. In the strong outbursts in 2008 October and in 2009 January and March, observed by Fermi gamma-ray burst monitor, the  period of 2.1s was clearly visible up to the energy $\sim 110$ keV (Kaneko et al., 2010).
The INTEGRAL detected pulsed soft gamma-rays from SGR/AXP 1E1547.0-5408 during its Jan-2009 outburst, in the energy band $20\div 150$ keV, showing a periodicity with P=2.1s (Kuiper et al. 2009).	
 This object is the only SGR/AXP with a relatively low period, all previous has periods exceeding $\sim 4s$.

  A low-magnetic-field  SGR0418+5729 was detected by Fermi gamma-ray burst detector (Rea et al. 2010). This soft gamma repeater with low magnetic field SGR0418+5729 was recently detected after it emitted bursts similar to those of magnetars. It was noted by Rea et al. (2010) that  "X-ray observations show that its dipolar magnetic field cannot be greater than
$7.5\cdot  10^{12}$ Gauss, well in the range of ordinary radio pulsars, implying that a high surface dipolar magnetic field is not necessarily required for magnetar-like activity".

\section{The Magnetar Model}

In the paper  of Duncan and Thompson (1992) was claimed, that dynamo mechanism in the new born rapidly rotating star may generate NS with a very strong  magnetic field $10^{14}\div 10^{15}$ G, called magnetars. These magnetars could be responsible for cosmological GRB, and may represent a plausible model for SGR. In the subsequent paper  (Duncan and Thompson 1995) the connection between magnetars and SGR was developed in more details.
The authors presented a  model for  SGRs, and the energetic 1979 March 5 burst, based on the existence of neutron stars with magnetic fields  much stronger than those of ordinary pulsars. They presented the following arguments point to a neutron star with B(dipole) $~5\cdot 10^{14}$ G as the source of the March 5 event (Duncan and Thompson 1995).

 1. Existence of such a strong magnetic field may spin down the star to 8 s period in the $~10^4$yr, what is the age of the surrounding supernova remnant N49.

 2. Magnetic field provide enough energy for the March 5 event.

 3. In presence such magnetic field  a large-scale interchange instability is developed with the growth time comparable to the ~0.2-s, close to the width of the initial hard transient phase of the March 5 event.

 4. A very strong magnetic field can confine the energy that was radiated in the soft tail of that burst.

 5.  A very strong magnetic field reduce the Compton scattering cross-section sufficiently to generate a radiative flux that is $\sim 10^4$ times the (non-magnetic) Eddington flux;

 6. The field decays significantly in $\sim 10^4 \div 10^5$ yr, as is required to explain the activity of soft gamma repeater sources on this time-scale; and

 7. The field power the quiescent X-ray emission $L_X \sim 7\cdot 10^{35}$ erg s$^{-1}$ observed by Einstein and ROSAT, as it diffuses the stellar interior. It is proposed that the 1979 March 5 event was triggered by a large-scale reconnection/interchange instability of the stellar magnetic field, and the soft repeat bursts are produces at cracking of the crust.

These suggestions were justified only by semi-qualitative estimations. Subsequent observations of $P$ and $\dot P$ in several SGR (McGill 2014), seems to support this model. However, when the rotation energy losses are much less than observed X-ray luminosity, $B$ estimations using $\dot P$ are not justified, because
magnetic stellar wind could be the main mechanism of angular momentum losses.
The jump in $\dot P$ observed in the giant burst of PSR1900+14 (Fig.\ref{fig8}) is plausibly explained by a corresponding increase of the magnetic stellar wind power, while the jump in the dipole magnetic field strength is hardly possible.
The jumps in $\dot P$, as well as in the pulse form (Figs.\ref{fig6},\ref{fig7}) have not been seen in the radio pulsars.
In the fall-back accretion model of SGR (Chatterjee et al. 2000, Alpar 2001, Tr\"umper et al. 2010, 2013) the estimations of the magnetic field using $P$ and $\dot P$ give the values characteristic for usual radiopulsars, when there is a  presence of a large scale magnetic field in the fall back accretion disk (Bisnovatyi-Kogan and Ikhsanov 2014).

When the energy density of the magnetic field is much larger that that of matter, as expected in the surface layers of the magnetar, the instability should be suppressed by magnetic forces.

The observations of radio pulsars, showing no traces of bursts, with magnetar magnetic fields and slow rotation (Section \ref{radio}), detection of SGR with a small rotational period and low magnetic field, estimated from $P$ and $\dot P$ values similar to radio pulsars (Section \ref{small}), gives a strong indication that inferred dipolar magnetic field strength and period cannot alone be responsible for the unusual high-energy properties of SGR/AXP. Therefore, another characteristic parameter should be responsible for a violent behaviour
of SGR/AXP. The unusually low mass of the neutron star was suggested by Bisnovatyi-Kogan (2012), Bisnovatyi-Kogan and Ikhsanov (2014) as a parameter, distinguishing SGR/AXP neuron stars from the majority of neutron stars in radio pulsars and close X-ray binaries.

\subsection{Angular momentum losses by a magnetized stellar wind}

A magnetic stellar wind carries away the stellar angular momentum $J$ as (Weber and Davis 1967)

\begin{equation}
\label{jwind}
\dot J_{wind}=\frac{2}{3}\dot M \Omega r_A^2,
\end{equation}
here $r_A$  is Alfven radius, where the energy density of the wind $E_w$ is equal to the magnetic energy density $E_B=B^2/(8\pi)$.
We consider the wind with a constant outflowing velocity $w$, which energy density is $E_w=0.5\rho w^2$. In a stationary wind
with a mass loss rate $\dot M$ the density is equal to

\begin{equation}
\label{dens}
\rho=\frac{\dot M}{4\pi w r^2}.
\end{equation}
For the dipole stellar field we have $B=\mu/r^3$, where $\mu=B_s r_*^3$ is the magnetic dipole moment of the star.
At the Alfven radius we have

\begin{equation}
\label{alf}
\rho_A=\frac{\dot M}{4\pi w r_A^2}, \quad E_{wA}=\frac{\dot M w}{8\pi r_A^2},\quad E_{BA}=\frac{\mu^2}{8\pi r_A^6}.
\end{equation}
From the definition of the Alfven radius $r_A$ we obtain its value as

\begin{equation}
\label{alf1}
E_{wA}=E_{BA},\quad r_A^4=\frac{\mu^2}{\dot M w}.
\end{equation}
The angular momentum of the star $J=I\Omega$, and when the wind losses (\ref{jwind}) are the most important, we obtain the
value of stellar magnetic field  as

\begin{equation}
\label{bw}
B^2_{wind}=\frac{9}{4}\frac{I^2\dot\Omega^2 w}{\Omega^2 \dot M r_*^6}.
\end{equation}
The angular momentum and energy losses by the dipole radiation which are main losses in
ordinary radiopulsars are written as (Pacini 1967)

 \begin{equation}
\label{psr}
L=\frac{B_s^2 \Omega^4 r_*^6}{c^3}, \quad \dot E=I\Omega\dot\Omega = L, \quad \dot J_{PRS}=\frac{L}{\Omega}.
\end{equation}
We obtain from (\ref{psr}) the magnetic field if the dipole radiation losses are the most important

\begin{equation}
\label{bpsr}
B^2_{PSR}=\frac{3I c^3\dot\Omega}{2\Omega^3 r_*^6}.
\end{equation}
The ratio of these two values is written as

\begin{equation}
\label{psrwind}
\frac{B^2_{PSR}}{B^2_{wind}}=\frac{2 c^3\dot M}{3I \Omega\dot\Omega w}=\frac{4}{3}\frac{\dot M w^2/2}{I \Omega\dot\Omega}\left(\frac{c}{w}\right)^3
=\frac{4}{3}\frac{F_{wind}}{\dot E_{rot}}\left(\frac{c}{w}\right)^3.
\end{equation}
Here $F_{wind}$ is the the energy flux carried away by the wind, and $\dot E_{rot}$ is rate of the loss of
rotational energy. For estimation of the energy flux carried away by the wind could be used the average
X and $\gamma$-ray luminosity of SGR/AXP $L_{x\gamma}$, and the wind velocity is of the order of the free fall velocity of the neutron star.
For low mass neutron star $M\leq 0.8 M_\odot$ we have $v_{ff}=\sqrt{\frac{2GM}{r_*}}\approx (c/3)$ at $M=0.6 M_\odot$,
$r_*=15$km, and

  \begin{equation}
\label{bb}
\frac{B^2_{PSR}}{B^2_{wind}}=36\frac{L_{x\gamma}}{\dot E_{rot}}.
\end{equation}
Using data from McGill(2014) and (\ref{bb}) we obtain for the magnetic fields of SGR  0526-66, SGR 1806-20, SGR 1900+14 the values
$10^{13}$, $1.7\cdot 10^{14}$, $6\cdot 10^{13}$ Gs respectively. While  the mechanical loss of the energy could exceed $L_{x\gamma}$,
these values of the magnetic field are suppose to be the upper limit if the magnetic field of these SGR.

\section{Model of nuclear explosion}
\label{nsnl}

It was shown by Bisnova\-tyi-Kogan and Chechetkin (1974), that in the neutron star crust full thermodynamic equilibrium is not reached, and a non-equilibrium layer is formed there during a neutron star cooling, see also Bisnovatyi-Kogan 2001.

 \begin{figure}[htp]
\centerline{\hbox{\includegraphics[width=0.9\textwidth]{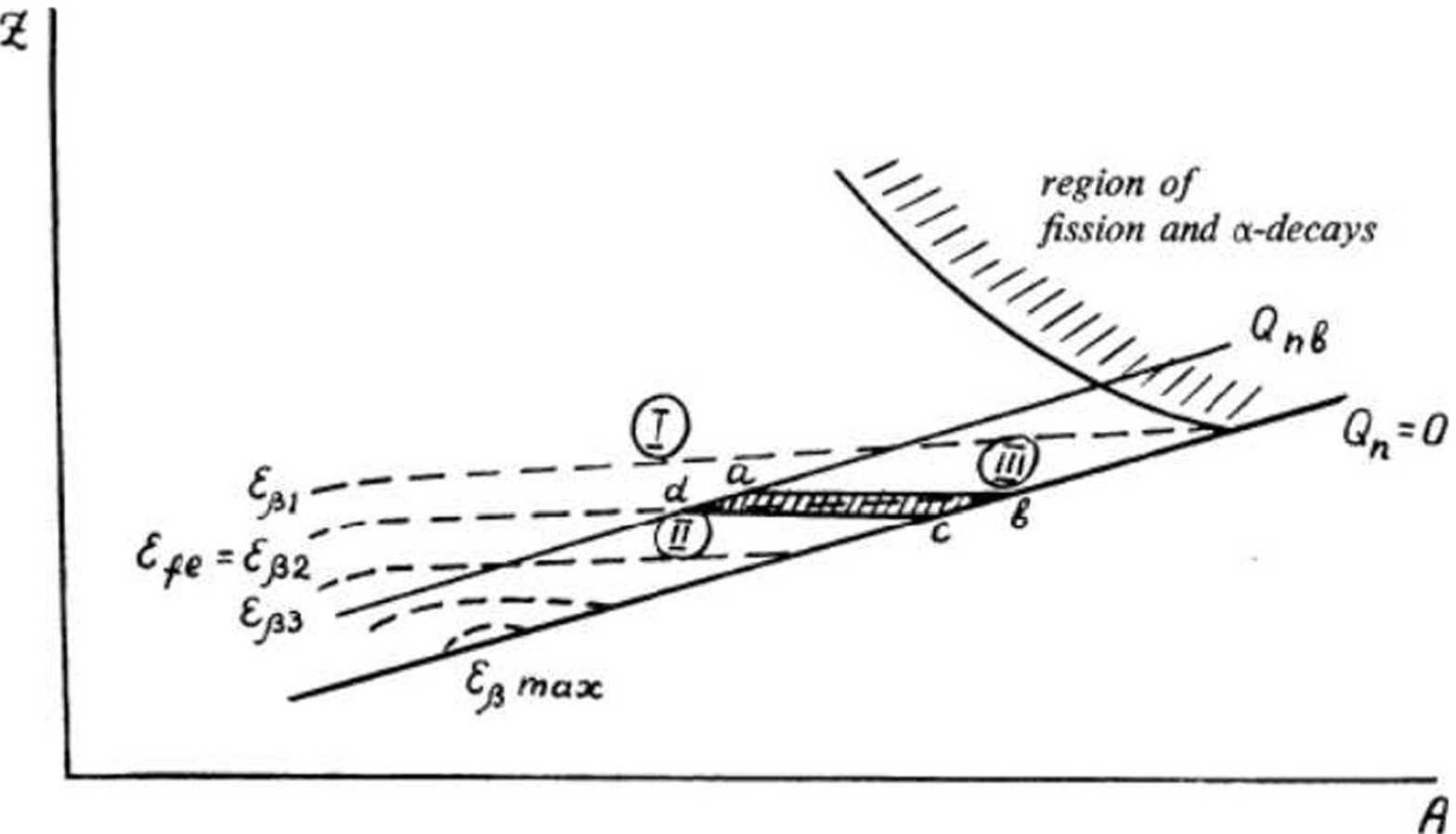}}}
\caption{ The formation of chemical composition at the stage of limiting equilibrium. The thick line $Q_n = 0$ defines the boundary of the region of existence of nuclei, the line $Q_{nb}$ separates region I, where photodisintegration of neutrons is impossible from regions II and III. The dashed lines indicate a level of constant $\varepsilon_\beta = Q_p - Q_n$; $\varepsilon_{\beta 1} < \varepsilon_{\beta 2}<  ... < \varepsilon_{\beta max}$. In region I we have $Q_n > Q_{nb}$; in region II we have $Q_n <  Q_{nb},\,\,\, \varepsilon_{fe} < \varepsilon_{\beta}$; and in region III we have $Q_n< Q_{nb},\,\,\, \varepsilon_{fe} > \varepsilon_{\beta}$. The line with the attached shading indicates a region of fission and $\alpha$-decay. The shaded region $abcd$ determines the boundaries for the values of (A,Z) with a limited equilibrium situation, at given values of $Q_{nb}(T)$ and $\varepsilon_{fe}(\rho)$, from Bisnovatyi-Kogan and Chechetkin (1974).}
\label{fig13}
\end{figure}

\noindent The non-equilibrium layer is formed in the region of densities and pressure $\rho_2<\rho<\rho_1$, $P_1<P<P_2$, with

$$\rho_1\simeq \mu_e 10^6 \left(\frac{8}{0.511}\right)^3\simeq 3.8\cdot 10^9\mu_e \,\,{\rm g/cm^3}\simeq 1.5\cdot 10^{10}\,\, {\rm g/cm^3}$$
$$\rho_2\simeq \mu_e 10^6 \left(\frac{33}{0.511}\right)^3\simeq 2.7\cdot 10^{11}\mu_e \,\,{\rm g/cm^3}\simeq  10^{12}\,\, {\rm g/cm^3}$$
$$P_1= 7.1 \cdot 10^{27}\,\, {\rm in\,\, cgs\,\, units}, \quad P_2 = 2.1 \cdot 10^{30}\,\,{\rm in\,\, cgs\,\, units}.$$

\begin{figure}[htp]
\centerline{\hbox{\includegraphics[width=1.2\textwidth]{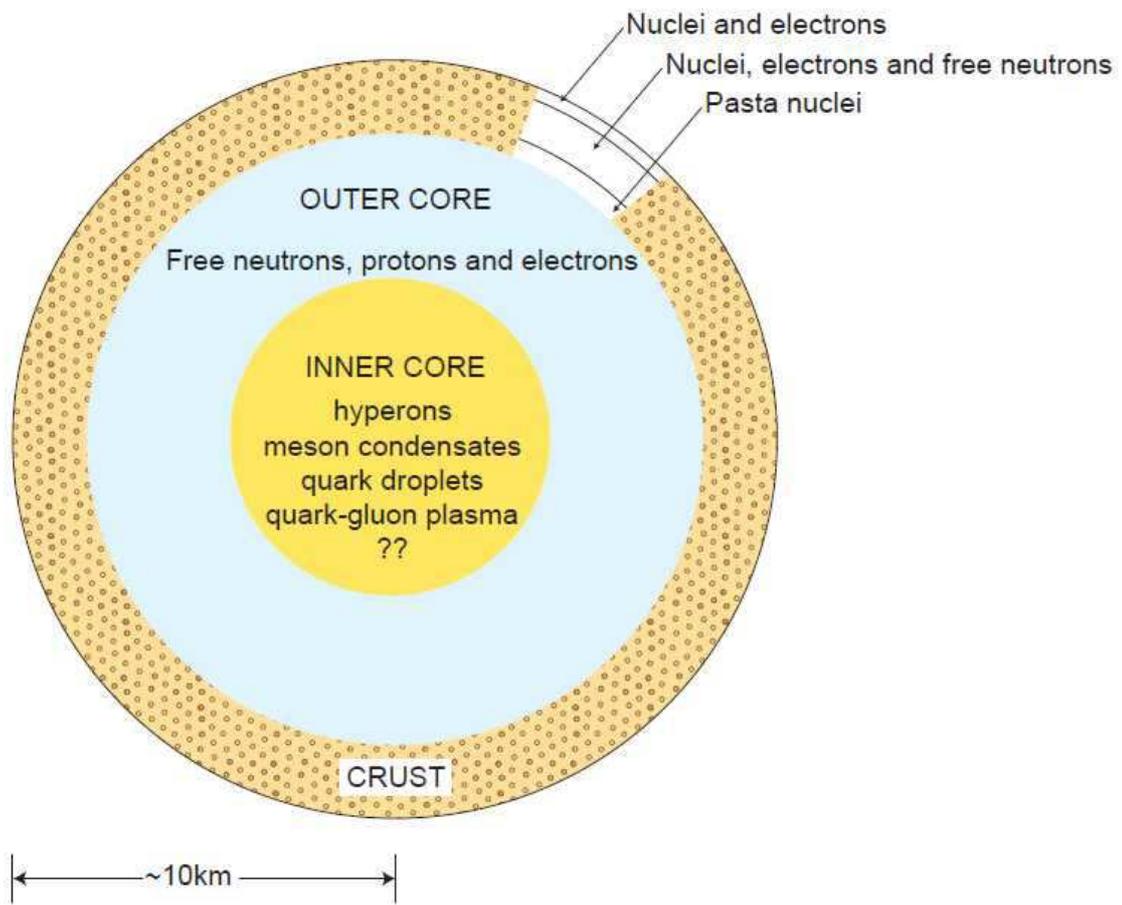}}}
\caption{ Schematic cross section of a neutron star, from Baym (2007).}
\label{fig10}
\end{figure}

\noindent The mass of the non-equilibrium layer is defined as (Bisnovatyi-Kogan and Chechetkin 1974)
$$M_{nl}=\frac{4\pi R^4}{GM}(P_2 - P_1) \simeq 0.1 (P_2-P_1)\simeq 2\cdot 10^{29}\,\, {\rm g}\,\,\simeq 10^{-4}\, M_\odot,$$

\noindent and the energy stored in this non-equilibrium layer is estimated as
$$ E_{nl}\simeq 4\cdot 10^{17} (P_2 - P_1) \approx 10^{48}\,\,{\rm erg}$$
Here a neutron star of a large $(\sim 2\, M_\odot)$ was considered, where the nonequilibrium layer is relatively thin, and its mass, and the energy store are estimated in the approximation of a flat layer. The nuclei in the non-equilibrium layer are overabundant with neutrons, so the number of nucleons per one electron is taken as $\mu_e\simeq 4$, and the energy release in the nuclear reaction of fission is about $5\cdot 10^{-3}\,c^2$ erg/g. A schematic cross-section of the neutron star is represented in Fig.\ref{fig10} from Baym (2007).

\begin{figure}[htp]
\centerline{\hbox{\includegraphics[width=1.2\textwidth]{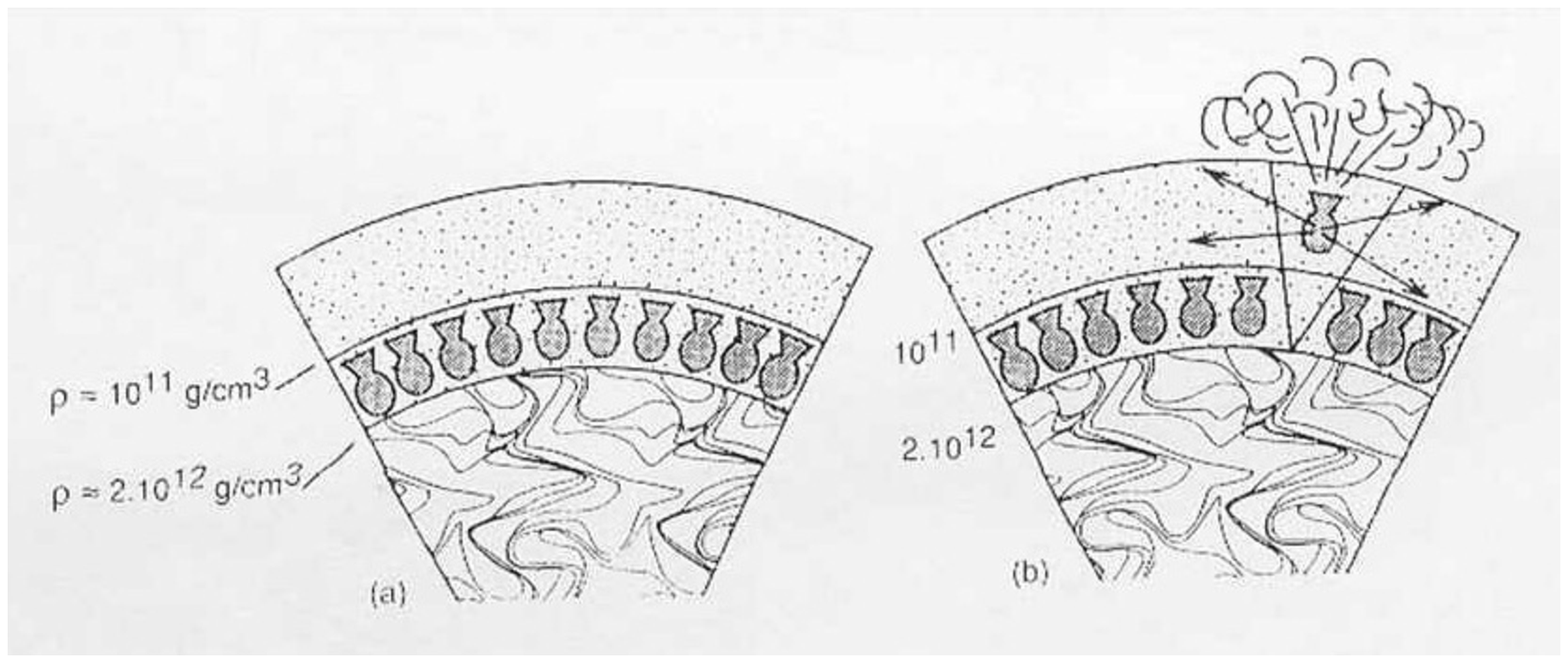}}}
\caption{The schematic picture of non-equilibrium layer in the neutron star: a) in a quiescent stage; b) after starquake and nuclear explosion, from Bisnovatyi-Kogan (1992).
\label{fig11}
}
\end{figure}

 \noindent Soon after discovery of gamma ray bursts the model of nuclear explosion was suggested (Bisnovatyi-Kogan et al. 1975), in which the non-equilibrium layer matter is brought to lower densities during a starquake. At the beginning GRB have been considered as objects inside the Galaxy, and the outburst was connected with period jumps in the neutron star rotation similar to those observed in the Crab nebula pulsar. It was suggested that: "Ejection of matter from the neutron stars may be related to the observed jumps of periods of pulsars. From the observed gain of kinetic energy of the filaments of the Crab Nebula ($\sim 2\cdot 10^{41}$ erg) the mass of the ejected material may be estimated as ($\sim 10^{21}$ g). This leads to energies of the $\gamma$-ray bursts of the order of $10^{38}-10^{39}$ erg, which agrees fully with observations at the mean distance up to the sources 0.25 kpc". A more detailed model of the strong 5 March 1979 burst, now classified as SGR 0526-66 in LMC,  was considered by  Bisnovatyi-Kogan and Chechetkin (1981). It was identified with an explosion on the NS inside the galactic disk, at a distance $\sim 100$ ps.
 The schematic picture of the nuclear explosion of the matter from the non-equilibrium layer is  presented in Fig.\ref{fig11}.

\noindent Cosmological origin of GRB, and identification of a group of non-stationary sources inside Galaxy as SGR/AXP lead to considerable revision of the older model, presented by Bisnovatyi-Kogan et al. (1975). It becomes clear that SGR represent a very rare and very special type of objects, which produce bursts much more powerful, than it was thought before from comparison with quakes in Crab nebula pulsar. Besides, the SGR are the only sources for which the nuclear explosions could be applied, because the energy release in the cosmological GRB highly exceed the energy store in the non-equilibrium layer.

\begin{figure}[htp]
\centerline{\hbox{\includegraphics[width=1.0\textwidth]{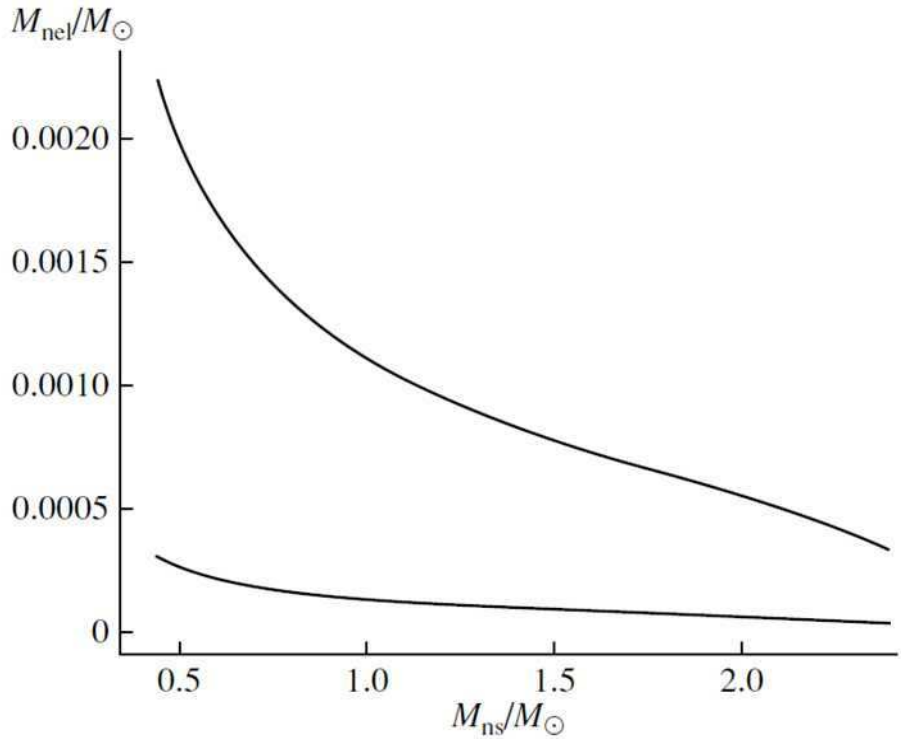}}}
\caption{
Dependence of the mass of the non-equilibrium layer on the neutron-star mass. The lines show the top and bottom boundaries of the layer mass measured from the stellar surface. The equation of state of the equilibrium matter \cite{bj,mbj}  was used to construct the model of the neutron star, with the boundaries of the layer specified by the densities. Using a non-equilibrium equation of state will increase the mass of the layer, but should not fundamentally change the values given in the figure, from Bisnovatyi-Kogan (2012), calculated and prepared by S.O. Tarasov.
}
\label{fig12}
\end{figure}

It was suggested by Bisnovatyi-Kogan (2012,2015), Bisnovatyi-Kogan and Ikh\-sa\-nov (2014), that the property, making the SGR neutron star so different from much more numerous of them in radio pulsars, single and binary X ray sources, is connected with the value of their mass, but not the magnetic field strength, see Camilo (2000), McLaughlin (2003), and Section \ref{radio}. Namely, it was suggested that the neutron stars in SRG/AXP have anomalously low mass, $(0.4 \div 0.8)M_\odot$, compared to the well measured masses in binary systems of two neutron stars, where neutron stars have masses $\ge 1.23\, M_\odot$ (Ferdman et al. 2014). The violent behaviour of the low-mass NS may be connected with much thicker and more massive non-equilibrium layer, and accretion from the fall-back highly magnetized accretion disk could trigger the instability, leading to outbursts explosions  (Bisnovatyi-Kogan and Ikhsanov 2014). The NS radius is increasing with mass rather slowly, so in a flat approximation the mass of non-equilibrium layer is inversely proportional to the mass. More accurate estimations have been obtained from calculations of neutron star models, presented in Fig.\ref{fig12}.
In Sect.\ref{nsnl} the calculated mass of the non-equilibrium layer $M_{nl}\approx 10^{-4} M_\odot$ was belonged to the neutron star with the mass $\sim 2\,M_\odot$ (see Bethe and Johnson 1975, Malone et al. 1975). For $M_{ns}=0.45\, M_\odot$ the mass of the non-equilibrium layer is $\sim 7$ times larger. The energy store reaches $\sim 10^{49}$ erg, what is enough for $\sim 1000$ giant bursts.

 The observational evidences for existence of neutron stars with masses, less than the Chandrasekhar white dwarf mass limit have been obtained by Janssen et al. (2008).
Observations of the binary pulsar system J1518-4904 indicated the masses of the components to be
$ m_p =0.72(+0.51, - 0.58 M_\odot)$,  $m_e = 2.00(+0.58, - 0.51) M _\odot$ with a $95.4$\% probability.
It was suggested by Bisnovatyi-Kogan and Ikhsanov (2014)  that low mass neutron stars could be formed in the scenario of the off-center explosion (Branch and Nomoto 1986), but more detailed
numerical investigation is needed to prove it. X-ray radiation of SGR/AXP in quiescent states was explained by Bisnovatyi-Kogan and Ikhsanov (2014) as a fall back accretion from the disk with a large scale poloidal magnetic field, what could also be a trigger for development of instability, leading to the mixing in the neutron star envelope, and nuclear explosion of the matter from the non-equilibrium layer.

\section{Conclusions}

\indent\indent 1. SGR are highly active, slowly rotating neutron stars.

2. Nonequilibrium layer (NL) is formed in the neutron star crust, during NS cooling, or during accretion onto it. It may be important for NS cooling, glitches, and explosions connected with SGR.

3. The mass and the energy store in NL increase rapidly with decreasing of NS mass.

4.  The properties of pulsar with high magnetic fields prove that inferred dipolar magnetic field strength and period cannot alone be responsible for the unusual high-energy properties of SGR/AXP.
The NL in low mass NS may be responsible for bursts and explosions in them.

5. The upper boundary of the magnetic fields in 3 most famous SGR, measured by the average $L_{x\gamma}$ luminosity is
about one order of magnitude lower than the values obtained using the pulsar-like energy losses of the rotational energy
of the neutron star.

6. Magnetar model of SGR, in which the energy of the observed bursts is provided by magnetic field annihilation, seems to be not relevant. Observations of quiet radiopulsars with a "magnetar" magnetic field, and of a low-field "magnetar", is the most important indication to that conclusion. A rapid growth of rotational periods, what is a favorite argument for a "magnetar" origin, is naturally explained by action of the magnetic stellar wind.
Besides, the high pressure of the magnetic field suppresses convection, which is needed in all annihilation models.

\bigskip

{\bf Acknowledgments}

\medskip

The work of  was partially supported by the Russian Foundation for Basic Research Grants No. 14-02-00728 and  OFI-M 14-29-06045, and the Russian Federation President Grant for Support of Leading Scientific Schools, Grant No. NSh-261.2014.2.

\end{document}